
\def\sqr#1#2{{\vcenter{\vbox{\hrule height.#2pt
      \hbox{\vrule width.#2pt height#1pt \kern#1pt
         \vrule width.#2pt}
      \hrule height.#2pt}}}}
\singlespace

\def\hg{\hat g}
\def\sqg{\sqrt{g}}
\def\sqhg{\sqrt{\hat g}}
\def\ab{{\alpha\beta}}
\def\glab{g_{\ab}}
\def\hglab{\hat g_{\ab}}
\def\guab{g^{\ab}}
\def\hguab{\hat g^{\ab}}
\def\pa{\partial}
\def\al{\alpha}
\def\be{\beta}
\def\si{\sigma}

\def\paal{\partial_\alpha}
\def\pabe{\partial_\beta}
 \PHYSREV
 \pubnum{6274}
 \date{June, 1993}
 \pubtype{T}
 \titlepage
 \title{ All-Order Quantum Gravity
         in Two Dimensions
         }
 \author{Thomas T. Burwick
 }
 \SLAC
 \abstract

 \singlespace
 We derive curvature counterterms in two-dimensional
 gravity coupled to conformal matter up to infinite order.
 By construction the higher-order action
 is equi\-valent to
 a finite first-order theory
 with auxiliary scalar.
 Due to this equivalence
 it shares the following remarkable properties:
 There is no need for gravitational dressing of
 the cosmological constant, quantization is consistent
 for any conformal anomaly $c$ of the coupled matter system,
 and if the coupled matter system is a $c=d~$-dimensional
 string theory in a Euclidean background then
 the effective string theory is $D=d+2~$-dimensional
 with Minkowski signature $(1,D-1)$.
 The resulting quantum theory favours flat geometries
 and suppresses both parabolic and hyperbolic
 singularities.
 \endpage
 \twelvepoint

 \chapter { Introduction }

 The quantum theory of gravity remains
 one of the deep mysteries of fundamental physics.
 Recent years have seen enormous effort to understand
 at least the case of a two-dimensional universe [1].
 There, at the classical level the Einstein-Hilbert action is
 trivial, but at the quantum level the Polyakov
 action shows up and causes the familiar difficulties
 if the system is coupled to matter [2].
 In the conformal gauge the Polyakov action turns into
 the Liou\-ville action. The quantum theory of this
 was first considered in [3,4] and a conformal field
 theory (CFT) treatment was given by David, Distler
 and Kawai [5].
 The occurrence of the Polyakov/Liouville action is
 most easily understood in the framework of
 path integrals by using the conformal gauge and
 going to a translation invariant
 measure [3,5-7].

 The central difficulty in quantizing two-dimensional
 gravity coupled to matter is the `Liouville
 problem'.
 It arises with
 the only non-trivial term of the two-dimen\-sional
 Einstein-Hilbert action: the cosmological constant.
 Coupling two-di\-men\-sio\-nal gravity
 to a conformal matter system of anomaly $c$,
 the matter system and gauge fixing contributions
 will induce the
 effective gravity action [3,2]
 $$
 S(g)={1\over \pi}\int d^2x\sqg\ \ (Q_0\ R\Delta^{-1} R
         + \lambda + \eta R)\ +\ S_C(g)
 \eqno\eq
 $$
 where $Q_0=(26-c)/24$, $R$ is the curvature in terms
 of the metric $\glab$,
 $\Delta={-1\over\sqg}\paal\sqg\guab\pabe$
 and $\lambda, \eta$ are arbitary constants.
 The first term is the Polyakov
 action (the $Q_0$ will be shifted after
 quantizing the metric),
 the next two terms give the Einstein-Hilbert action.
 If the system is defined on a closed manifold of genus $h$
 then ${1\over 2\pi}\int d^2x\sqg R = 2(1-h)$,
 the Euler number.
 (We use the conventions of [7].)
 The $S_C$ is supposed to collect possible counterterms.
 Using the conformal gauge
 $\glab = e^{2\sigma}\hglab$
 where
 $\hglab$ is a background-metric,
 quantizing the Liouville-mode $\si$
 and applying
 CFT methods to (1.1) without counterterms,
 the cosmological
 constant
 $\sqg \lambda = \sqhg e^{2\si} \lambda$
 is seen
 to be not of weight (1,1) [5]. Therefore the action (1.1)
 is not
 a scalar, and general covariance is broken at the
 quantum level unless appropriate counterterms $S_C$
 are found.
 This is the `Liouville problem'.

 The usual consequence is to replace the cosmological
 constant
 by
 a `dressed' version which is of
 weight (1,1) [5]
 $$
 e^{2\si}\quad\rightarrow\quad e^{2\kappa\si}
 \qquad , \qquad
 \kappa = {1\over 12}~\bigl(~25~-~c~-~
 \sqrt{(1-c)(25-c)}\bigr)~.
 \eqno\eq
 $$
 This approach, however, has major drawbacks:
 the theory loses its geometrical character,
 i.e., the renormalized
 action can no longer be formulated in terms of the
 metric $g_\ab$
 if $\kappa\neq 1$ and the dressed cosmological constant
 has to be complex if $1<c<25$
 so that the renormalization (1.2) becomes
 senseless in this region.

 Given these difficulties it is rather
 natural to search for an alternative
 renormalization program:
 the inclusion of higher-order counterterms.
 Such a renor\-ma\-li\-zation procedure will be the content
 of this paper.
 Instead of (1.2) we will arrive at counterterms of the
 form
 $$
 S_C(g) = {Q\over 2\pi\alpha} \int d^2x\sqg\ \sum_{m=2}^{\infty}
                  q_m\ (\alpha R)^m
 \eqno\eq
 $$
 where $Q=(24-c)/24, \al\sim (length)^2$ is some renormalization
 scale and $q_m$ are the coefficients that will be determined
 in what follows.
 We will find the renormalization $Q_0\rightarrow Q$ in (1.1)
 after quantizing the gravity sector.
 Notice that in (1.3) no terms show up
 with derivatives acting on the curvature.
 This will be a feature of our result but may
 have been expected from the beginning by
 carefully reexamining the principle of ultralocality
 stated by Polchinski [8,7].
 (We come back to this point in the summary.)

 Many attempts have been made to avoid
 the problems associated with (1.2):
 W-gravity, quantum groups and others.
 It is fair to say that none of them
 succeeded so far.
 Recently, however, it was examined in
 greater detail that another problematic feature
 of two-dimensional gravity, the instabilities
 of surfaces related to covering with spikes
 and branched polymers is cured at scales
 $\ll 1/\mu,~\mu\sim (length)^{-2}$
 if a $(1/\mu)R^2$-term is added to (1.1) [9].
 This is easily understood by realizing that
 the quadratic term suppresses arbitrary high
 curvature.
 In [9] the $R^2$-theory was studied by
 introducing an auxiliary field $\phi$ and
 analyzing instead the
 action $\int d^2x\sqg (\phi R + \mu\phi^2)$.
 Combining this with the results of [10]
 it can be seen immediately that the $R^2$-term
 is even enough to assure that the cosmological
 constant is of weight (1,1) without the shift
 (1.2) (see (2.7) and (2.8) below).
 However, new problems arise with
 the $\phi^2$-term.
 In the conformal gauge $\sqrt{\hat g}e^{2\si}\phi^2$
 will not be of weight (1,1).
 This is why the success of [9] would be
 complete only if $\mu\rightarrow 0$.
 This limit, however, establishes the
 constraint of vanishing curvature which
 may conflict with the Gauss-Bonnet
 theorem and therefore requires
 additional renormalization [10,11].
 The potential $\phi^2$ is ill-defined.
 Instead we have to work with another
 potential. This will be our starting point
 in the following.
 In summary, we take the point of view that
 the partial success of the $R^2$-theory
 is due to including a quadratic curvature term
 that is already enough to avoid (1.2) and
 suppresses arbitrary high curvature,
 the failure however is due to neglecting
 terms of even higher order.

 Apparently a quantum analysis with (1.3) is
 a tough problem. Fortunately, there is a short cut.
 Let us write the counterterms in (1.1) as
 (we work with Euclidean signature)
 $$
 e^{-S_C(g)} = \int D_g\phi\ e^{-\widetilde S_C(g,\phi)}
 \eqno\eq
 $$
 This will allow us to use recent progress made
 in understanding two-dimensional gravity in the
 presence of an additional scalar $\phi$,
 a `dilaton'. In particular, we may use that if
 the metric-dilaton action is taken to be
 $$
 \widetilde S_C(g,\phi)\ =\  {1\over 2\pi}
     \int d^2x\sqrt{g}\
      (\ Q\phi R +\mu  e^\phi\ )
 \eqno\eq
 $$
 then the Liouville problem is absent,
 (1.1) with (1.4) describes a well-defined
 quantum theory,
 and quantization is possible for any conformal
 anomaly $c$ of the coupled matter system [10].
 Contrary to the $R^2$-theory that
 corresponds to a $\phi^2$-potential [9], the exponential
 potential in (1.5) describes a consistent
 quantum theory.
 Many aspects of
 the remarkable success
 related to (1.5)
 have reappeared in other
 frameworks including black hole physics [12,13].
 (Using conformal gauge the action (1.1) turns into
 a non-linear sigma model, see (2.10) below.
 The metric-dilaton theories of [13] can be written
 in the same form. However, contrary to the discussion
 here, the relation between $X^\pm$, the Liouville
 mode $\si$, and the dilaton in [13] is such that $X^\pm$
 are bounded and therefore the quantum theories in [13]
 are incomplete.)
 A recent study in terms of non-linear sigma models
 may be found in [14].

 The dilaton $\phi$ appearing in (1.5) is
 an auxiliary field.
 It is therefore natural to eliminate it and ask
 for the form of the counterterms $S_C(g)$
 that have to be included in (1.1).
 Doing so will be the content of this paper.
 Indeed, we find that the integration (1.4) can
 be performed explicitly.
 The metric-scalar theory (1.5) may then be understood
 as a first-order formulation
 of the higher-order theory with $S_C(g)$ and (1.1)
 may be
 seen as the effective action obtained from integrating
 out the auxiliary field.

 In section 2 we review the
 cornerstones of the CFT analysis that reveal why
 (1.4) with (1.5) leads to a consistent quantum theory
 such that there is no need for the gravitational
 dressing (1.2).
 We also argue for the finiteness of (1.1)
 with (1.4), (1.5) from a diagrammatic point
 of view.
 In section 3 we compute the higher-order
 counterterms $S_C(g)$ by explicitly
 integrating out the auxiliary field in (1.4).
 Performing a path-integration requires some
 regularization of the surface. We work with standard
 triangulations. In section 4
 we comment on the
 surprisingly (and convincingly!) sensible
 geometrical content of the resulting higher-order
 theory. Section 5 contains our summary.

 \chapter { The Gravity Action with
                Auxiliary Field }

 Let us use (1.4) to express $S_C(g)$
 in terms of the first-order action (1.5).
 We will now shortly review that (1.4) with (1.5)
 indeed provides a solution of
 the Liouville problem.
 A detailed discussion can be found in [10].
 We will also give a simple diagrammatic argument
 for the finiteness of the quantum theory.
 Notice that although the scalar in (1.5) is
 an auxiliary field it will obtain a mixed
 kinetic term with the Liouville mode
 in the conformal gauge. This gauge will
 be used in this section.

 Upon quantizing the metric,
 the Liouville measure together with the
 dilaton-measure in (1.4) induces a
 renormalization of the coefficient in (1.1):
 $$
 Q_0\quad\rightarrow\quad Q\ =\ {{24-c}\over 24}
 \eqno\eq
 $$
 In the conformal gauge $\glab=e^{2\si}\hglab$
 the curvature is
 $R=e^{-2\si}(\widehat R+\widehat\Delta\si)$
 and so the effective gravity action (1.1) with (2.1),
 (1.4) and (1.5) reads
 $$
 \eqalign{
 S(\sigma,\phi) = {Q\over \pi} \int d^2x\sqhg\
  (\ \hguab&\paal\si\pabe\si + 2\si \widehat R \cr
  &+ {1\over 2}\hguab\paal\phi\pabe\si
   + {1\over 2}\phi\widehat R+ V(\si, \phi))}
 \eqno\eq
 $$
 with potential
 $$
 V(\si, \phi) = \lambda e^{2\si}\ +\ \mu e^{2\si+\phi}~.
 \eqno\eq
 $$
 In (2.2) we rescaled $\lambda, \mu$ and did not
 write pure background terms.
 Using the local gauge $\hg_\ab =\delta_\ab$
 the stress-energy
 tensor $\widehat T_\ab =
 {-4\pi\over\sqhg}{\delta S\over\delta\hguab}$
 is
 $$
 \eqalign{
 \widehat T_\ab\ =\ &2Q(\delta_\ab(\hat{\pa}^\gamma\si\pa_\gamma\si
   + {1\over 2}\hat{\pa}^\gamma\phi\pa_\gamma\si + V(\si, \phi))\cr
 &- 2\paal\si\pabe\si + 2\paal\pabe\si
 - 2\delta_\ab\hat{\pa}^2\si\cr
 &+ {1\over 2}(\paal\pabe\phi - \delta_\ab\hat{\pa}^2\phi
 - \paal\phi\pabe\si - \pabe\phi\paal\si))
 }
 \eqno\eq
 $$

 The reason for writing $S_C(g)$ in its first-order
 form (1.4) with (1.5) is that the
 effective action (1.1) turns into (2.2)
 which can be analyzed by using simple
 CFT methods (for an easy review of these see [15]).
 Applying the standard David, Distler and Kawai
 procedure [5], we treat the potential
 terms (2.3) as a perturbation. Then (2.2)
 without (2.3) is a CFT and, using
 complex coordinates,
 the $\si$ and $\phi$-fields obtain
 mixed propagators
 $$
 \eqalign{
 <\phi(z,\bar z)\phi(w,\bar w)>
            &= {2\over Q} \ln|z-w|^2\cr
 <\phi(z,\bar z)\si(w,\bar w)>
            &= {-1\over 2Q}\ln|z-w|^2\cr
 <\si(z,\bar z)\si(w,\bar w)>   &= 0\cr
 }
 \eqno\eq
 $$
 For the analytic part of (2.4)
 the operator product expansion (OPE) gives
 $$
 :\widehat T(z):\ :\widehat T(w):\ =\ {{1\over 2}(26-c)\over (z-w)^4}
    + \Bigl[{2\over (z-w)^2}
    + {\partial_w\over z-w}\Bigr]\ :\widehat T(w): + ...
 \eqno\eq
 $$
 The complete system is free of the conformal
 anomaly since the anomaly $(26-c)$ arising
 in (2.6) just cancels the anomaly coming from
 the matter and gauge fixing sector.

 The crucial test for consistency is whether
 the terms (2.3) can be included without breaking
 general coordinate invariance.
 This is where (1.1) without counterterms fails
 and eventually leads to (1.2). However,
 in the presence of (1.4) with (1.5),
 the propagators (2.5) imply that
 $$
 :\widehat T(z):\ :e^{2\al\si(w) + \be\phi(w)}: =
      \Bigl[{1\over (z-w)^2}
    + {\partial_w\over z-w}\Bigr]\ :e^{2\al\si(w) + \be\phi(w)}:
 \eqno\eq
 $$
 has indeed solutions with $\al=1$:
 $$
 (\al,\be)\quad =\quad (1,0)\quad \hbox{or}\quad (1,1)~.
 \eqno\eq
 $$
 The first solution corresponds to the
 cosmological constant in (1.1), the second to
 the scalar self-coupling in (1.5).
 Thus the terms in (2.3) are of weight (1,1),
 they do {\it not} violate
 general covariance at the quantum level,
 there is {\it no} need for gravitational dressing
 and {\it no} restriction on the
 matter anomaly $c$ appears!

 In this paper, we are approaching a higher-order
 theory of gravity. As a consequence the
 metric propagator may include unphysical modes
 due to higher-order derivatives:
 $$
 {1\over\pa^4}\ =\ {1\over{2i\epsilon}}\Biggl(
   {1\over{\pa^2 - i\epsilon}} -
   {1\over{\pa^2 + i\epsilon}} \Biggr)\quad,\quad
   \epsilon\rightarrow 0
 \eqno\eq
 $$
 A natural question to worry about is whether these
 will induce a ghost problem.
 Again this question is most easily studied in the
 first-order formulation using (2.2).
 With $X^+ = \sqrt{Q}~(2\si+\phi),~
 X^- = \sqrt{Q}~2\si$ the action
 (2.2) can be written
 $$
 \eqalign{
 S(\si,\phi)\ =\ {1\over{2\pi}}\int d^2x\sqhg\Bigl(
    {1\over 2}\hat\pa^{\al} X^+&\paal X^-
     + \sqrt{Q}(X^++X^-)\widehat R\cr
    &+ 2\lambda Q e^{X^-/\sqrt{Q}}
     + 2\mu Q    e^{X^+/\sqrt{Q}\ }\Bigr)~.\cr
 }\eqno\eq
 $$
 If the coupled matter system
 is a c-dimensional string theory in a Euclidean
 background the
 gravitational part (2.10)
 with $X^\pm=X^0\pm X^{c+1}$
 turns this into
 a D=c+2-dimen\-sio\-nal
 string theory in a background with
 Minkowski signature $(1,D-1)$.
 The negative contributions related to (2.9)
 cause no problems.
 The propagators of (2.2), (2.10) are given by
 (2.5)
 and are well-defined without violating causality.

 Notice that supersymmetrization
 is straightforward.
 To eliminate tachyons
 appearing in a string theory like (2.10)
 the action (1.5) may be replaced by its
 supersymmetric version using the results
 of [16].

 The finiteness of (2.2), (2.10) may also
 be understood in a diagrammatic way.
 Whenever we work with a two-dimensional field-theory
 with second-derivative kinetic term and polynomial
 potential, a diagram with $P$ internal
 lines and $V$ vertices will have the
 superficial degree of divergency
 $$
 \omega~=~2~(P-[V-1])~-~2~P~=~2~(1-V)~.
 \eqno\eq
 $$
 Only diagrams with $V=1$ (or with $V=1$ subdiagrams)
 can be divergent (see fig. 1).
 Such diagrams will appear if
 no $S_C(g)$ is included in (1.1). Then in the
 conformal gauge the cosmological constant together
 with the $<\si\si>$-propagator
 will induce such divergent diagrams.
 It is therefore not surprising that
 a renormalization (1.2) may occur.
 What happens if an $R^2$-term is included?
 Such a term will have an enormous impact
 due to higher derivatives like (2.9)
 contributing to the metric propagator.
 Again the analysis may be simplified by using an
 auxiliary field $\phi$. In the conformal
 gauge the $R^2$-term will then
 result in an action like (2.10)
 with the replacement
 $$
 e^{X^+/\sqrt{Q}}
 \qquad\rightarrow\qquad
    Q^{-1}~e^{X^-/\sqrt{Q}}~(X^+-X^-)^2
 \eqno\eq
 $$
 The propagators are given by (2.5).
 Now the Liouville propagator vanishes
 and in terms
 of $X^\pm$ the only non-zero propagator
 is $<X^+X^->$. It is then immediately obvious
 that the cosmological constant no longer
 causes a problem: using the
 vertices $\lambda e^{X^-/\sqrt{Q}}$ no
 divergent diagram can be drawn.
 This is the reason for the first solution in (2.8).
 On the other hand there is a new problem arising
 with (2.12).
 The quadratic $\phi$-potential introduces vertices with
 both $X^+$ and $X^-$ legs. Closing these
 with $<X^+X^->$ gives loop-diagrams
 of the divergent type.
 Therefore
 additional potential terms have to
 cancel these diagrams:
 $$
 \sum_{n=0}^\infty ~{Q^{-n/2}\over n!}~
 e^{X^-/\sqrt{Q}}~(X^+-X^-)^n~
   =~e^{X^+/\sqrt{Q}}~.
 \eqno\eq
 $$
 With (2.13) no divergent loop-diagram can be
 constructed and no additional counterterms are
 needed!
 This is again the potential of (2.10)
 and the reason why
 the action has to be of the form (1.5).
 The potential (2.13) corresponds
 to the second solution in (2.8).
 For $S_C(g)$ this implies that terms
 of order higher than $R^2$ have to be included.
 These terms will be derived
 in the next section.

 Before calculating the integral (1.4)
 it may be worthwhile to
 remember the equi\-valence
 between higher-order gravity and
 first-order formulations with
 an additional scalar $\phi$
 at the classical level.
 A higher-order gravity theory, given by some
 arbitrary function of the curvature
 $$
    {1\over 2\pi\al} \int d^2x \sqrt{g} \ f(R)
 \eqno\eq
 $$
 where $\al\sim (length)^2$,
 can be turned into a first-order theory provided
 that the definition
 $$
  \phi = {1\over\al}\ f'(R)
 \eqno\eq
 $$
 with $'={d\over dR}$ allows to solve for $R$ in terms
 of $\phi$. Then a potential can be defined by
 $$
 \ V(\phi) =  {1\over\al}\bigl(\ f(R) - R f'(R)\ \bigr)
 \eqno\eq
 $$
 and (2.14) turns out to be equivalent to the first-order
 system
 $$
  {1\over 2\pi}\ \int d^2x \sqrt{g}\ ( \phi R + V(\phi) )
 \eqno\eq
 $$
 Therefore a higher-order gravity theory
 can be formulated
 as a first-order system by adding a scalar.
 This was first observed by Higgs [17],
 later rediscovered
 by Whitt [18] for $f(R) = R + \alpha R^2$
 in $D=4$ (of course
 the above procedure is possible for any
 space-time dimension)
 and has been extended to higher powers of
 Ricci and Riemann
 tensors in [19].
 A recent application is [20] where (2.17) has
 been used to
 study four-dimensional black hole solutions in
 higher-derivative gravity.

 Our first-order action (1.5) is indeed
 of the form (2.17).
 Assuming $\mu \sim 1/\al$ and neglecting topological
 terms, its classical analog (2.14) is given by
 $$
 f(R)~=~Q~\al R~\ln~(\al R)~.
 \eqno\eq
 $$
 This action does not seem to make sense,
 in particular around zero curvature.
 The equivalence between (2.14) and  (2.17)
 is however only a
 classical one. At the quantum level the corresponding
 equivalence
 must be established by integrating out the dilaton.
 Only if $V(\phi) \sim \phi^2$ will the result of this
 integration agree with the classical procedure.
 We have seen that
 in two dimensions $V(\phi)$ has to be exponential
 and
 so there is no alternative to performing the integration.
 This integration will be subject of the next
 section and will result in
 a higher-order quantum action
 clearly more acceptable
 than (2.18).

 \chapter { The Higher-Order Gravity Action }

 We now return to (1.4) and integrate out
 the auxiliary field.
 This will lead to the explicit form of the
 counterterms $S_C(g)$.
 In the last section we saw that
 using the action (1.5) in (1.4) ensures that
 we arrive at a consistent quantum theory (1.1)
 with no need for gravitational dressing (1.2).
 The absence of a kinetic term for $\phi$
 in (1.5) is linked to the
 propagators (2.5)
 and thus essential for the quantum
 consistency reflected in the OPEs (2.6), (2.7).
 As a consequence
 the dilaton $\phi$ is an auxiliary field,
 there is no damping of
 arbitrary high frequencies
 and it is natural to eliminate it.
 Any path integration requires some
 regularization. We will work with
 triangulations (for an introduction
 to these see [21]).

 We regularize the surface by a triangulation
 with V vertices, E edges and F faces (fig.2).
 The triangles are assumed to be of
 area $\pi\al$, where $\al\sim (length)^2$ is
 the regulator that will enter in (1.3).
 The total area A of the surface is
 $$
 A=\int d^2x\sqrt{g} \simeq F\cdot \pi\al
 \eqno\eq
 $$
 The $\alpha$ is a kind of UV-cutoff. The
 continuum limit is obtained
 from $\al\rightarrow 0, F\rightarrow\infty$.
 The basic identities relating V, E and F are
 $V-E+F=2(1-h)$ and $2E=3F$.
 The regularization is then established by the
 replacements
 $$
 \eqalign{
 \int d^2x\sqrt{g(x)}\ (...)&\simeq\
    \sum_{i=1}^V\ s_i\ (...)
    \quad ,\quad s_i\ =\ {N_i\over 3}\ \pi\al\cr
 R(x)\ &\simeq\ R_i\ =\ {{2\pi}\over s_i}\
     \Bigl( 1-{s_i\over{2\pi\al}} \Bigr)\cr
 }\eqno\eq
 $$
 where $N_i$ counts the nearest neighbours of
 vertex $i$.
 Moreover, we have to regulate the measure in the
 path-integral. With (3.2) we get for scalar fields $X$
 with value $X_i$ at vertex $i$
 $$
 \parallel\delta X\parallel^2\ \simeq\
     \sum_{i=1}^V s_i\ (\delta X_i)^2~.
 \eqno\eq
 $$
 Using (3.2), (3.3)
 we may write (1.4) as
 $$
 e^{-S_C(g)}\simeq \prod_{i=1}^V\sqrt{s_i}
    \ \int_{C} d\phi_i \
    \ e^{-\widetilde S_C(s_i,\phi_i)}
 \ =\ \prod_{i=1}^V\sqrt{s_i}\ e^{-S_C(i)}
 \eqno\eq
 $$
 where
 $$
 \widetilde S_C(s_i,\phi_i) =
     {s_i\over {2\pi}}
     \Bigl( Q\phi_iR_i + \mu e^{\phi_i}\Bigr)~.
 \eqno\eq
 $$
 The path-integral factorizes which will be essential
 for performing the integration.
 We see immediately that no
 terms like $\al^3 R\Delta R,...$ with $\Delta$
 acting on $R$
 will appear and the counterterms will indeed be
 of the form (1.3).

 We have to specify the integration contour $C$.
 To integrate along the real axis seems to be senseless.
 This is obvious
 after recognizing the structure
 of a $\Gamma$-function:
 $$
 \int_{-\infty}^{+\infty}d\phi\ e^{a\phi\ -\ be^{\phi}}
 \quad =\quad e^{-a\ln b}\ \Gamma(a)\quad ,
 \quad Re~a>0 \quad (b>0)~.
 \eqno\eq
 $$
 This integral is divergent
 for Re $a\leq 0$. Applied to (3.4)
 this would imply that whenever a given geometry
 has one vertex with  $QR_i\geq 0$
 the $\phi$-integration would be divergent.

 One should not try to establish convergence
 by changing the integrand in (3.6), simply because
 (1.5), (3.5) is the only action we know to
 solve the problems of quantization.
 This will exclude subtracting the singularities
 from the integrand leaving as finite pieces the
 $\Gamma$-function between its poles.
 Apparently, the only way we can make sense out
 of (3.4) is to choose the integration
 contour to be different from (3.6).
 This, however has to be subject to another
 requirement: we may change the contour only such that
 perturbation theory, in particular
 the OPEs (2.6), (2.7) and the arguments
 leading to (2.11) and below
 are not affected.
 Perturbation theory is obtained by coupling the
 fields to sources and then performing a (real) shift
 in the path-integration. Therefore, we should
 not cut the $\phi$-integration in (3.7),
 for example
 to $[0,\infty)$ or $(-\infty,0]$.
 Cutting the contour would introduce boundaries in
 field space which are subtle issues to
 deal with.

 Fortunately, there is a way to make sense out
 of (3.4) such that all these requirements are satisfied.
 Instead of cutting the contour we may extend it.
 Let us consider the complex $e^\phi$-plane.
 The integration in (3.6) is along the positive
 real axis. For
 non-integer $a$ (or $Q{s_i\over 2\pi}R_i$ in (3.4))
 there is a cut in this plane
 starting from the origin.
 Instead of using the $\phi$-integration of (3.6) we
 may go to the integration contour shown in fig. 3a, b:
 $$
 \int_{-\infty}^{+\infty}d\phi
  \quad\rightarrow \quad
  {1\over {2\pi i}} \int_{C} d\phi
 \eqno\eq
 $$
 In the $-e^\phi$-plane the contour comes in along the
 positive real axis, encircles the origin
 counterclockwise and goes back to infinity.
 Using this replacement in (3.6) we obtain [23]
 $$
 {1\over{2\pi i}}\int_{C} d\phi\
   e^{a\phi - be^\phi} = e^{-a\ln (-b)}
   \ {1\over{\Gamma(1-a)}}
   \quad ,\quad |a|~< \infty \quad (b<0)
 \eqno\eq
 $$
 Here, we have chosen the phase such that the
 integral is real.
 In (3.8) we arrived at Hankel's representation
 of the
 $\Gamma$-function
 which is well-defined every\-where
 on the complex plane.
 (A replacement similar to (3.7) has been used in
 Liouville theory
 in the context of regularizing the area-integration
 to obtain finite expression for correlation
 functions [22]).
 The finiteness of (3.8) may be understood by
 realizing that for Re $a>0$ the vertical part of $C$
 gives vanishing integration.
 For Re $a\leq 0$ this integration diverges, thereby
 cancelling corresponding divergencies arising in the
 integration along the horizontal parts of $C$.
 Obviously perturbation theory is not affected
 if $\phi$ takes values along this contour.

 Using the integration contour $C$ in (3.4) we may
 apply (3.8) and obtain
 $$
 S_C(i)\ =\
  - Q{s_i\over{2\pi}}R_i
  \ln \Bigl(-\mu{s_i\over 2\pi}\Bigr) +
  \ln {\Gamma(1+Q{s_i\over{2\pi}}R_i)}
 \eqno\eq
 $$
 In order to write this as an action, we should remember
 the heat-kernel expansion for the $\delta$-function
 which is at the origin of the Liouville action appearing
 in the conformal gauge [7,6]. This reads
 $$
 {\Delta^2x\sqrt{g}\over2\pi}\bigl(\ R(x) +
 {1\over\Lambda^2}\ \bigr) = 6\pi\Delta^2x
 \ \delta^{(2)}(0)
 \eqno\eq
 $$
 where $\Lambda\sim (length)\rightarrow 0$
 is some UV-cutoff. Using our regularization the
 analog relation is
 $$
 {s_i\over2\pi}\bigl(\ R_i +
 {1\over\al}\ \bigr) = 1
 \eqno\eq
 $$
 which is easily derived from (3.2).
 Including $\sum_{i}$ in (3.9) we may use (3.11)
 to write $S_C$ as an action. According to (3.2)
 this action is equivalent to
 $$
 \eqalign{ S_C(g)\ =
 \ {1\over {2\pi\al}}\int d^2x\sqrt{g}
   \Biggl(Q\al R&\ln(1+\al R)\cr
   &+(1+\al R)\ln \Gamma
     \bigl(1+Q{{\al R}\over{1+\al R}}\bigr)\Biggr)~.\cr
 }\eqno\eq
 $$
 This is the action we have been looking for.
 Although it may look rather complicated
 we will see below that it is just of a form
 that is geometrically meaningful.
 Notice that instead of writing the
 term ${-1\over24\pi}\int d^2x\sqg\ R\Delta^{-1}R$
 in (3.12) which is induced from the dilaton measure
 in (1.4), we keep track of it by renormalizing
 the coefficient $Q_0$ in (1.1).
 Quantizing the Liouville mode also, this
 will imply
 $$
 Q_0\quad\rightarrow\quad Q\ =\ {{24-c}\over 24}
 $$
 as mentioned before in (2.1).
 In the discrete case this metric dependence
 of the measure shows up in the
 triangulation itself and the
 factor $\sqrt{s_i}$ appearing in (3.4).
 Such a factor comes with any matter field measure.
 In (3.14) we separated it from (3.9).
 In (3.12) we also did not write the topological
 term $-2Q(1-h)\ln(-\mu\al)$.
 The coefficient $\mu$
 does affect only this topological term.
 A natural choice would be $-\mu\sim 1/\al$.
 (Notice that the $\phi$ along
 the horizontal part of $D$ has imaginary part $\pm i\pi$
 so that with $\mu<0$ in (1.5) and (3.5) $\mu e^\phi>0$.)

 Contrary to (2.18) the terms (3.12) are well-behaved
 around zero curvature,
 they just vanish if $R=0$.
 The action (3.12) is formulated at some
 scale $\alpha$.
 As a consequence possible fluctuations
 of the geometry are restricted.
 Indeed, with (3.2) we always have
 $$
 {(\ 1+\al R_i\ )\ =\ {6\over N_i}~ >~ 0}~.\qquad
 \eqno\eq
 $$
 Therefore no singularity problems arise in (3.12).

 In (1.3) we expected an infinite series of
 counterterms:
 $$
 S_C(g)~= {Q\over 2\pi\al}
  \int d^2x\sqg~\sum_{m=2}^{\infty}~
  q_m~(\al R)^m~.
 \eqno\eq
 $$
 Such a form may be more familiar from perturbation
 theory. So let us deduce the coefficients $q_m$
 from (3.12). We have to use
 $$
 \eqalign{
 \ln (1+z)~&=~-~\sum_{n=1}^{\infty}~
      {1\over n}~(-z)^n\cr
 \Bigl( {1\over{1+z}}\Bigr)^k~&=~
  k~\sum_{n=0}^{\infty}{1\over {k+n}}
   {k+n\choose k}~(-z)^n\cr
 }\eqno\eq
 $$
 if $|z|<1$, and [23]
 $$
 \ln\Gamma(1+w)~=~-\gamma w~+~\sum_{n=2}^\infty~
  {\zeta(n)\over n}~(-w)^n
 \eqno\eq
 $$
 if $|w|<1$, where $\gamma =0.57721...$ is Euler's constant
 and $\zeta$ is Riemann's Zeta-function.
 With $z=\al R, w=\al R/(1+\al R)$ this can be
 applied to (3.12) and the coefficients in (3.14)
 turn out to be
 $$
 q_m~=~{(-1)^m\over{m-1}}\Biggl[~1~+~
   \sum_{k=1}^{m-1}~kQ^k~{m-1\choose k}~
   {\zeta(k+1)\over{k+1}}\Biggr]~.
 \eqno\eq
 $$
 Upon expanding (3.12) there is also a
 topological term $-2\gamma Q(1-h)$ arising.
 We did not write this in (3.14).
 Since the radius of convergence for this series
 is at $\alpha R_i = (1+Q)^{-1}$ one may prefer to work
 with the closed expression (3.12) instead of expanding it.

 \chapter { The Geometrical Implications }

 In the foregoing sections we determined $S_C(g)$ by insisting
 on quantum consistency in the continuum limit.
 Let us now study the geometrical implications
 of the resulting quantum action.
 We mentioned in the introduction that already the
 presence of a quadratic curvature
 term $R^2$ in (1.1) is
 enough to pacify the problem of spikes
 covering the surface and branched polymers
 at small scales [9].
 At these scales the surfaces turn out to be
 smooth since high curvature is suppressed by
 the quadratic term.
 There is, however, no reason to expect
 that such a feature would survive the inclusion
 of even higher-order terms.
 In this section we will find
 that although our $S_C(g)$ contains
 curvature terms up to infinite order it is indeed
 of a form that favours flat geometries.

 Since $S_C(g)\simeq \sum_i{S_C(i)}$ the effect
 of (3.12) is to introduce a weight
 $$
 e^{-S_C(i)}\ =
   \Bigl({N_i\over 6}\Bigr)^{Q\bigl[1-{N_i\over 6}\bigr]}
   {1\over \Gamma \Bigl(1+Q\bigl[ 1-{N_i\over 6} \bigr] \Bigr)}
   ~e^{-2\eta'(1-{N_i\over 6})}
 \eqno\eq
 $$
 at each vertex $i$.
 The dependence on the conformal dimension $c$
 of the coupled matter system enters
 via $Q=(26-D)/24$ where $D=c+2$ is
 the dimension of the effective theory
 (see (2.10)).
 To arrive at (4.1) we used (3.9) without
 the topological term that was dropped
 in (3.12). The last factor in (4.1)
 corresponds to an
 additional $\eta'{1\over\pi}\int d^2x\sqrt{g}R$.
 With $\eta'=Q\gamma/2$
 this will cancel the topological term arising
 when (3.12) is expanded into (3.14).
 Notice that the first factor in (4.1) was
 studied some time ago [24].
 Here the essential new feature is the appearence of
 the inverse $\Gamma$-factor.

 Let us first discuss the region $D<26~(Q>0)$.
 The weight (4.1) is zero for
 $$
 N_i~=~0~\quad \hbox{or}
   \quad 6(1+Q^{-1}),~6(1+2Q^{-1}),~...
 \eqno\eq
 $$
 The maxima between these zeros are obtained if $N_i$
 obeys
 $$
 Q({6\over N_i}-1) - Q\ln{N_i\over 6}
   + Q \psi(1+Q(1-{N_i\over 6})) + 2\eta'~=~0
 \eqno\eq
 $$
 where $\psi(z)={d\over dz}\ln{\Gamma(z)}$.
 To begin we may concentrate on the region between
 the first two zeros (see fig. 4).
 We comment on the other regions when discussing
 singularities.
 Using $\eta'=Q\gamma/2$
 the maximum of (4.1) is then obtained at
 $$
 N_i~=~6\quad :\quad {d\over dN_i}~e^{-S_C(i)}~=~0
 \eqno\eq
 $$
 since $\psi(1)=-\gamma$.
 Therefore our $S_C(g)$ is such that it favours
 flat geometries.

 Before discussing singularities
 we should recall some basics  about Riemann surfaces.
 For any point on the surface the geometry
 of a local neighbourhood may be thought of as
 being induced from a three-dimensional Euclidean space
 into which this local region is embedded [25].
 In order to regularize the quantum theory we represented
 the Riemann surfaces in terms of $F$ equilateral
 triangles of area $\pi\alpha$.
 At a vertex $i$ with $N_i$ nearest
 neighbours (see fig. 2) the local geometry will be
 flat ($N_i=6$), parabolic ($N_i<6$), or
 hyperbolic ($N_i>6$).
 It is easily visualized that
 in a three-dimensional embedding space
 a locally hyperbolic geometry (`saddle-surface') gets
 singular at $N_i=12$.
 A global embedding may require a higher-dimensional
 space [25].
 At finite scale $\al$
 global embedding theorems have to be applied and
 vertices with $N_i\geq 12$ may occur.

 What about singularities? Let us first study the
 case of an
 effective $D=2$ dimensional string theory,
 i.e., the case of pure gravity.
 Then $Q=1$ and the region between the first two zeros
 of (4.2) is $0 < N_i < 12$ (fig. 4).
 Obviously the weight (4.1) decreases for
 small $N_i$-values, thereby suppressing parabolic
 singularities. On the other hand we just mentioned
 that also $N_i=12$ has geometrical significance.
 It corresponds to a hyperbolic singularity in the
 continuum limit.
 At finite scale $\al$ we have to consider also
 vertices with $N_i\geq 12$. It turns out that
 vertices with arbitrary large $N_i$ are strongly
 suppressed (see fig. 5).
 This is a remarkable result.
 We find that our $S_C(g)$ not only
 favours flat geometries, but also
 suppresses both parabolic and hyperbolic
 singularities.

 When matter is included, $D>2 ~(Q<1)$,
 the geometric content gets less obvious.
 The maximum
 of (4.1) remains at flat geometries.
 With increasing $D$
 the suppression of non-flat geometries is weakened.
 In particular the second zero $6(1+Q^{-1})$ in (4.2)
 (and all higher zeros) will increase (see fig. 4).
 Given the interpretation of $D$ as target-space dimension
 this may reflect that with increasing embedding
 dimension there is `more space' to include triangles.
 Notice that random triangulations will also include
 configurations that do not correspond to Riemann surfaces.
 Nevertheless, vertices with arbitrary large $N_i$
 are strongly suppressed also for $D>2$
 (see fig. 5) and Riemannian geometries
 may be expected to be dominant
 even if arbitrary triangulations are considered.
 Definite statements should be left to
 future studies, e.g., in the framework of
 numerical simulations.

 Alternatively, the region beyond the second
 zero, $N_i\geq 6(1+Q^{-1})$, can be eliminated completely
 if the  weight (4.1) is defined in terms of
 the expansion (3.14), (3.17) instead of (3.12).
 Then whenever $N_i\geq 6(1+Q^{-1})$ the series
 is divergent and the weight is zero.

 To study the critical limit $D\rightarrow 26$
 and $D>26$
 it is helpful to look back
 at the first-order fomulation (2.2), (2.10).
 Let us choose the
 coupled matter system to be a $c=d$-dimensional
 string theory in a flat background
 with Euclidean metric $\delta_{ij}$.
 With $X^\pm=X^0\pm X^{d+1}$ and not writing the
 potential terms (2.3) the gravitational
 action (2.2), (2.10) gives
 $$
 \eqalign{
 S(\si,\phi)~+ {1\over 4\pi}\int d^2x&\sqhg~
   ~\hguab\paal X^i\pabe X^j\delta_{ij}\cr
   &=~{1\over 4\pi}\int d^2x\sqhg~
   (~\hguab\paal X^\mu\pabe X^\nu\eta_{\mu\nu}
   ~+~O(Q)~)\cr
 }\eqno\eq
 $$
 where $\eta_{\mu\nu}=~$diag$(-1,+1,...,+1),~
 \mu,\nu = 0,1,...,D-1=d+1$ is a Minkowski metric.
 As usual the target-space metric and
 world-sheet metric in (4.5) are related by
 $$
 \hglab~=~{\partial X^\mu \over \pa x^\al}
    {\partial X^\nu \over \pa x^\be}~\eta_{\mu\nu}
    ~+~O(Q)~.
 \eqno\eq
 $$
 The background geometry on the world-sheet
 is induced from the target-space and corrected
 by quantum contributions for the non-critical case.
 With (4.5), (4.6) the relation to the critical theory
 is obvious
 since $O(Q)$ vanishes as $D\rightarrow 26$.

 For pure Liouville theory the region $D>26$
 is known to be troublesome. There $Q<0$, and
 the kinetic Liouville term will cause
 the $\si$-integration to diverge. This is usually
 dealt with by rotating $\si\rightarrow i\si$,
 a transformation with unclear geometrical content.
 In our case we have another problem. With $Q<0$
 the weight (4.1) and in particular the zeros (4.2)
 do not seem to make sense. Both problems may be cured
 by rotating the two world-sheet
 coordinates $x^\al\rightarrow ix^\al$
 (followed by $X^j\rightarrow iX^j$ to
 leave the matter sector invariant).
 As a result the gravity action will change sign
 which may be expressed by $Q\rightarrow -Q$.
 This is just the correction we need.
 There may indeed be some motivation for
 this transformation.
 If we move $D$ from $D<26$ to $D>26$
 the $X^\pm$-values are rotated from the real
 axis to the imaginary axis. This is due to
 the factor $\sqrt{(26-D)/24}$ included in their
 definition. Since
 target-space and world-sheet coordinates $X^\mu$
 and $x^\al$ are related
 by (4.5), (4.6)
 it may then be natural to rotate also the $x^\al$
 into the imaginary direction.
 Since we work with Euclidean signature on
 the world-sheet this procedure simply corresponds
 to applying a different Wick-rotation for $D>26$.
 Notice that even with $Q\rightarrow -Q$ the
 second zero $6(1-Q^{-1})$ of (4.1)
 will be at $N_i<12$ if $D>50$.
 It is unclear whether this procedure
 is appropriate. Of course, one should not
 exclude that a sensible quantization can only
 be established for $D\leq 26$.
 However, since from the
 perturbative point of view the quantum theory is
 well-defined for any $D$ (see section 2)
 there ought to be hope that also in the
 non-perturbative setting we could make sense
 out of the theory if $D>26$. Again, we should
 leave definite statements to future investigations.
 However, one may speculate that the behaviour of
 (4.1) in the presence of matter $(Q<1)$ is related
 to geometrical phenomena that become dominant
 in the region $D>26$. Thus it may be advisable
 to obtain a clear geometrical picture for
 the region $D<26$ first.

 Finally, we may consider the
 scaling behavior of the partition function.
 Let us rescale the area $A$
 by $A~\rightarrow ~\Lambda A$.
 Using the first-order
 formulation (1.4), (1.5) where the
 conformal gauge turns (1.1) into (2.2),
 the rescaling of the area is expressed
 by $2\si\rightarrow 2\si+\ln\Lambda$.
 We may use the standard method to turn the sum
 over geometries into a sum over partition
 functions for fixed
 area $Z(A)$ [27,5]. The
 string susceptibility $\xi$ is then defined by [26]
 $$
 Z(A)~\sim~e^{-{\lambda\over \pi} A}~A^{\xi -3}~.
 \eqno\eq
 $$
 Compensating in (2.2) the shift of $\si$ by
 shifting
 the $\phi$-integration
 $\phi\rightarrow \phi-\ln\Lambda$
 one obtains $\xi = 2+{D-26\over 12}(1-h)$.
 This value does not agree
 with the semi-classical
 value $\xi^0\approx {D\over 6}(1-h),
 ~D\rightarrow -\infty$ [27,28].
 If on the other hand we set $\mu\sim 1/\al$
 and accompany the rescaling of the area (3.1)
 by $\al\rightarrow \Lambda\al$
 then the Lagrangian (1.5) is invariant,
 no shift in $\phi$ is necessary and one
 obtains $\xi=2+{D-26\over 6}(1-h)$. This result would agree
 with the semi-classical limit.
 To clarify the situation
 notice that the weight (4.1)
 is independent of the area $\al$.
 In (3.9) we found that upon regularizing
 (1.4), (1.5) in terms of triangles and integrating
 out $\phi$, the only $\al$-dependence shows
 up in a topological term $-2Q(1-h)\ln(-\mu\al )$.
 This term was canceled in (4.1).
 With $A\rightarrow \Lambda A,~\al\rightarrow \Lambda \al$
 it would disturb the semi-classical limit
 unless $\mu$ scales as $1/\al$.
 Therefore, assuming that $\mu$ in (2.2)
 scales with $1/\al$ corresponds
 to cancelling the topological factor
 arising in the integration (1.4).
 It is another appealing feature of our results that,
 contrary to the string suscep\-ti\-bi\-li\-ty resul\-ting
 from (1.2) [2,5], our $\xi$
 is real for any $D$.

 \chapter { Summary}

 The (Liouville-)problem of quantizing two-dimensional
 gravity with conformal matter of anomaly $c$
 can be solved if the metric $g_\ab$
 is accompanied by an auxiliary
 field $\phi$.
 Choosing the appropriate action (1.5) for $\phi$
 the cosmological constant turns out to
 be of conformal weight (1,1) without any need
 for gravitational dressing (1.2) and quantization becomes
 possible for any $c$ [10].
 (See also related work in [12-14].)
 This
 can be shown by using standard David, Distler
 and Kawai CFT-methods.
 In section 2 we also
 gave simple diagrammatic arguments.
 (It is convenient to analyze the quantum consistency
 in the conformal gauge $g_\ab=e^{2\si}\hat g_\ab$.
 In this gauge $\phi$ obtains a mixed propagator with the
 Liouville-mode $\si$ and if the coupled matter system
 is a d-dimensional
 string theory in Euclidean background the effective
 theory is a $D=d+2$ string theory in a background with
 Minkowski-signature $(1,D-1)$.)

 The auxiliary character of $\phi$ is essential
 for quantum consistency. Any kinetic term for $\phi$
 would destroy the finiteness of the theory.
 Given this auxiliary nature of $\phi$ we wanted
 to eliminate it and asked for the corresponding
 higher-order action $S_C(g)$.
 This was done in section 3.
 The $\phi$-potential has to be exponential and
 so $\phi$ could be eliminated only by
 explicitly performing its path-integration.
 (Only with quadratic potential would the
 path-integration agree with simple elimination
 by equation of motion.)
 In order to regularize this integration
 we triangulated the two-dimensional universe
 and had to extend the $\phi$-contour to the complex
 plane.
 Then $\phi$ could be integrated out explicitly
 and, using the triangulated version of the
 heat-kernel expansion, the result could be
 formulated as an action with higher-order curvature
 terms.

 The weights (4.1) introduced by the higher-order
 terms are such that flat geometries are dominating.
 Discussing the case of pure gravity
 in section 4, we found both
 parabolic and hyperbolic singularities to be suppressed.
 In the presence of matter
 the geometrical content becomes less obvious.
 Still flat geometries are favoured but
 the suppression of non-flat geometries is weakened
 as the critical limit ($D=26$) is approached.
 Without a clear geometric understanding of the
 region $D<26$ it may be too early to make
 statements on $D>26$.
 In particular we do not know whether our result
 should also be applied to the non-Riemannian
 regime, e.g., to triangulations that do not
 correspond to regular surfaces. We have to leave
 definite statements
 to further investigations, possibly in the
 framework of numerical simulations.

 Essential features of the
 results presented in this paper may have been
 expected from the beginning by using
 Polchinski's principle of ultralocality [8,7].
 In two dimensions the Einstein-Hilbert action is
 trivial. The only dynamics is induced from the measure
 in the path-integral.
 Any counterterms will have to make this measure
 well-defined and so the path-integration.
 Therefore, one may speculate that counterterms $S_C(g)$
 correspond to ambiguities in the measure.
 The principle of ultralocality states that
 the measure and thus any ambiguity
 is a local product over
 space-time points.
 This is often used to argue that any ambiguity
 in the measure is restricted to the cosmological
 constant. However, the notion of `locality'
 can only be defined together with a regularization
 of the path-integral.
 Here, we regularized the two-dimensional surface
 in terms of triangles and therefore `locality' refers to
 the plaquette around each vertex $i$, see fig. 2.
 Local quantities are quantities defined on each plaquette
 seperately.
 If, following the principle of ultralocality,
 we assume the  counterterms to reflect the local
 structure of the measure, then
 we may include curvature
 terms $R_i$ but not
 differences $R_i-R_j,i\neq j$, or in terms of
 the scalar field no $\phi_i-\phi_j,i\neq j$.
 It is therefore not surprising that the counterterms
 factorize as it happens in (4.1).
 So we find the auxiliary nature of $\phi$
 and in $S_C(g)$ the absence of terms with derivatives
 acting on $R$ related to structures of the measure.

 The theory presented certainly has a
 rich structure that is waiting to be discovered.
 It should open the avenue to study string theories
 with $c>1$ (which actually is $c>0$ in
 our framework).
 Future studies should illuminate the geometrical
 content in more detail. Correlation functions
 may be most conveniently discussed along the lines
 of [29,11] using the first-order formalism.
 This will also hold for analyzing the spectrum.
 Some matrix-model formulation would be highly wellcome.
 Here, we introduced triangulations only as a tool
 to regularize the two-dimensional surface and
 integrate out the auxiliary field to gain some
 insight into the higher-order theory.
 Using this
 regularization the next step would be to implement
 the summation over geometries as a summation over
 triangulations.
 Given the particularly simple form of the
 first-order formulation with auxiliary field
 one may speculate that a two-matrix model
 would be appropriate.
 (Notice in the conformal gauge the
 action (2.10) includes coupling to a
 background charge and some progress has been
 made to relate minimal CFTs to two-matrix
 models [30].)
 As long as matrix-model formulations are
 absent one may study non-perturbative effects
 using numerical simulations.
 These should be straightforward
 using the weight (4.1).
 Dynamical triangulations with higher-order
 curvature terms have recently been studied in [31]
 (with lattice sizes up to 400,000 triangles)
 and earlier in [24].
 Contrary to the weights used there, the weight (4.1)
 originates from a consistent continuum theory.

 As a conclusion we find that to cure the Liouville
 problem and quantize two-dimensional gravity with
 conformal matter of arbitrary anomaly $c$ there is a
 choice of two prices that have to be paid:
 Either an auxiliary field $\phi$ is included or
 - equivalently and discussed in this paper -
 curvature terms up to infinite order will appear.

 \vskip1.0truecm
 {\bf Acknowledgements} It is a pleasure to thank
 A.H. Chamseddine, J. Garc\'{\i}a-\-Bellido,
 M. Peskin and L. Susskind
 for valuable discussions. This work was supported in part
 by the Swiss National Science Foundation (SNF)
 and in part by the U.S. Department of Energy,
 contract DE-AC03-76SF00515.

 \endpage

 \centerline{\bf REFERENCES}
 \vskip1.0truecm
 \item{[1]}  E. D'Hoker, {\it Mod. Phys. Lett.}
             {\bf A6} (1991) 745; \hfil\break
             N. Seiberg, {\it Prog. Theor. Phys.
             Suppl.} {\bf 102} (1990) 319.

 \item{[2]}  A. M. Polyakov,
    {\it Mod. Phys. Lett.} {\bf A2} (1987) 893;
    \hfil\break
    V.G. Knizhnik, A. Polyakov and A. Zamolodchikov,
    \hfil\break
    {\it Mod. Phys. Lett.} {\bf A3} (1988) 819.

 \item{[3]}  A. M. Polyakov,
    {\it Phys. Lett.} {\bf B103} (1981) 207.

 \item{[4]}  T. L. Curtwright and C. B. Thorn,
    {\it Phys. Rev. Lett.} {\bf D48} (1982) 1309;
    \hfil\break
    E. Braaten, T. Curthwright and C. Thorn,
    {\it Phys. Lett.} {\bf B118} (1982) 115,
    {\it Ann. Phys.(NY)} {\bf 147} (1983) 365;
    \hfil\break
    J. L. Gervais and A. Neveu,
    {\it Nucl. Phys. } {\bf B199} (1982) 59.

 \item{[5]}  F. David,
    {\it Mod. Phys. Lett.} {\bf A3} (1988) 1651;
    \hfil\break
    J. Distler and H. Kawai,
    {\it Nucl. Phys. } {\bf B321} (1989) 509.

 \item{[6]}  N. Mavromatos and J. Miramontes,
    {\it Mod. Phys. Lett.} {\bf A4} (1989) 1847;
    \hfil\break
    E. D'Hoker and P. S. Kurzepa,
    {\it Mod. Phys. Lett.} {\bf A5} (1990) 1411.

 \item{[7]} E. D'Hoker and D.H. Phong,
    {\it Rev. Mod. Phys.} {\bf 60} (1988) 917,
    and references therein.

 \item{[8]} J. Polchinski,
    {\it Comm. Math. Phys.} {\bf 104} (1986) 37.

 \item{[9]} H. Kawai and R. Nakayama,
    {\it Phys. Lett.} {\bf B306} (1993) 224.

 \item{[10]}  A.H. Chamseddine,
    {\it Phys. Lett.} {\bf B256} (1991) 379;
    \hfil\break
    {\it Nucl. Phys. } {\bf B368} (1992) 98;
    \hfil\break
    T.T. Burwick and A.H. Chamseddine,
    {\it Nucl. Phys.} {\bf B384} (1992) 411.

 \item {[11]} S. F\"orste,
   {\it Z. Phys.} {\bf C55} (1992) 525;
   {\it Z. Phys.} {\bf C57} (1992) 663.

 \item {[12]}
    S.D. Odintsov and I.L. Shapiro,
    {\it Phys. Lett.} {\bf B263} (1991) 183;
    \hfil\break
    {\it Mod. Phys. Lett.} {\bf A7} (1992) 437;
    \hfil\break
    J.G. Russo and A.A. Tseytlin,
    {\it Nucl. Phys.} {\bf B382} (1992) 259.

 \item {[13]}
    S.P. de Alwis,
    {\it Phys. Lett.} {\bf B289} (1992) 278;
    {\it Phys. Lett.} {\bf B300} (1993) 330;
    \hfil\break
    Bilal and C.Callan,
    {\it Nucl. Phys. } {\bf B394} (1993) 73;
    \hfil\break
    J.G. Russo, L. Susskind, and L. Thorlacius,
    {\it Phys. Rev.} {\bf D46} (1992) 3444;
    {\it Phys. Rev.} {\bf D47} (1993) 533.

 \item {[14]} A.A. Tseytlin,
    {\it Phys. Rev.} {\bf D47} (1993) 3421.

 \item {[15]} D. L\"ust and S. Theisen,
    `Lectures on String Theory',
    Springer Verlag Berlin, 1989.

 \item {[16]} A.H. Chamseddine,
   {\it Phys. Lett.} {\bf B258} (1991) 97.

 \item {[17]} P.W. Higgs,
    {\it Nuovo Cimento} {\bf 11} (1959) 816.

 \item {[18]} B. Whitt,
   {\it Phys. Lett.} {\bf B145} (1984) 176.

 \item {[19]} G. Magnano, M. Ferraris,
                and M. Francaviglia,
   \hfil\break
   {\it Gen. Relativ. Gravit.} {\bf 19} (1987) 465;
   {\it Class. Quantum Grav.} {\bf 7} (1990) 557;
   \hfil\break
   A. Jakubiec and J. Kijowski,
   {\it Phys. Rev.} {\bf D37} (1988) 1406;
   \hfil\break
   J.D. Barrow and S. Cotsakis,
   {\it Phys. Lett.} {\bf B214} (1988) 515.

 \item {[20]} S. Mignemi and D.L. Wiltshire,
    {\it Phys. Rev.} {\bf D46} (1992) 1475.

 \item {[21]} P. Ginsparg,
   `Matrix Models of 2d Gravity',
   Lectures given at Trieste Summer School,
   Italy, July 1991, LA-UR-91-9999,
   \hfil\break
   Bulletin Board: hep-th@xxx.lanl.gov - 9112013.

 \item {[22]} I.I. Kogan,
   {\it Phys. Lett.} {\bf B265} (1991) 269.

 \item {[23]} I.S. Gradshteyn and I.M. Ryzhik,
   `Table of Integrals, Series, and Products',
   Academic Press, 1980.

 \item {[24]} J. Ambj\o rn, B. Durhuus and J. Fr\"ohlich,
   {\it Nucl. Phys.} {\bf B275[FS17]} (1986) 161;
   \hfil\break
   D.V. Boulatov, V.A. Kazakov, I.K. Kostov
   and A.A. Migdal,
   {\it Nucl. Phys.} {B275[FS17]} (1986) 641;
   \hfil\break
   A. Billoire and F. David,
   {\it Nucl. Phys.} {\bf B275[FS17]} (1986) 617;
   \hfil\break
   J. Jurkiewicz, A. Krzywicki and B. Peterson,
   {\it Phys. Lett.} {\bf 168B} (1986) 273;
   \hfil\break
   F. David,
   {\it Nucl. Phys.} {\bf B257[FS14]} (1985) 45.

 \item {[25]} M. Gromov and I.Rhokhlin,
    `Embeddings and Immersions
    in Riemannian geometry',
    {\it Russ. Math. Surveys} {\bf 25} (1970) 1.

 \item {[26]} D. Weingarten,
   {\it Phys. Lett.} {\bf B90} (1980) 280.
   \hfil\break
   T. Eguchi and H. Kawai,
   {\it Phys. Lett.} {\bf B114} (1982) 247.

 \item {[27]} A.B. Zamolodchikov,
   {\it Phys. Lett.} {\bf B117} (1982) 87.

 \item {[28]} S. Chaudhuri, H. Kawai and S.-H. Tye,
   {\it Phys. Rev.} {\bf D36} (1986) 1148.

 \item {[29]} A.M. Polyakov,
   {\it Mod. Phys. Lett.} {\bf A6} (1991) 635;
   \hfil\break
   M. Goulian and M. Li,
   {\it Phys. Rev. Lett.} {\bf 66} (1991) 2051;
   \hfil\break
   H. Dorn and H.J. Otto,
   {\it Phys. Lett.} {\bf B232} (1989), 327;
   {\it Phys. Lett.} {\bf B280} (1992), 204.

 \item {[30]} J.-M. Daul, V.A. Kazakov and I.K. Kostov,
   `Rational Theories of 2d Gravity from Two-Matrix
   Models', CERN-TH-6834/93, March 1993;
   \hfil\break
   M. Douglas, in `Random Surfaces and Quantum Gravity',
   Carg\`ese 1990, eds. O. Alvarez et al., Plenum,
   1991, p.77.

 \item {[31]} N.Tsuda and T.Yukawa,
   {\it Phys. Lett.} {\bf B305} (1993) 223.

 \endpage
 \endpage
 \centerline{\bf FIGURE CAPTION}
 \vskip1.0truecm

 \item{\rm FIG.1}
 In two-dimensional field theories with
 second-derivative kinetic term and
 polynomial potential only one-vertex (sub-)diagrams
 can be divergent.

 \item{\rm FIG.2}
 A piece of triangulation of a two-dimensional surface.
 The broken lines show the dual graph, describing
 plaquettes of area $s_i=N_i\pi\al/3$ around a vertex $i$
 with $N_i$ nearest neighbours.

 \item{\rm FIG.3}
 Integration contour for the auxiliary field $\phi$.

 \item{\rm FIG.4}
 The weight (4.1) between its first two zeros,
 introduced at each vertex $i$ due to
 the higher-order terms.

 \item{\rm FIG.5}
 The weight (4.1) suppresses arbitrary large
 numbers $N_i$ of nearest neighbours.
 (The solid and broken curves correspond to fig. 4.)

 \bye